\documentclass[12pt]{article}

\usepackage{graphics}

\begin{document}

\baselineskip=16pt 

\title{Voltage-biased I-V characteristics in the multi-Josephson junction 
	model of high T$_c$ superconductors}

\author{\normalsize 
        Shoichi SAKAMOTO and Hideki MATSUMOTO\\ 
        Department of Applied Physics, Seikei University\\  
 	Kichijoji Kitamachi 3-3-1\\ 
        Musashino-shi, 180-8633 JAPAN \\ 
	and\\ 
	Tomio KOYAMA\\ 
	Institute for Materials Research, Tohoku University\\ 
	Katahira 2-1-1, Sendai, 980 JAPAN\\ 
        and\\ 
	Masahiko MACHIDA\\ 
	Center for Promotion of Computational 
	Science and Engineering\\ 
	Japan Atomic Research Institute\\ 
	2-2-54 Nakameguro, Meguro-ku, Tokyo, 153 JAPAN
	}
\date{\today}
\maketitle
\begin{abstract}
	By use of the multi-Josephson junction model, we investigate 
voltage-biased I-V characteristics. Differently from the case of the single 
junction, I-V characteristics show a complicated behavior due to inter-layer 
couplings among superconducting phase differences mediated by the charging 
effect. We show that there exist three characteristic regions, which are 
identified by jumps and cusps in the I-V curve. In the low voltage region, 
the total current is periodic with trigonometric functional increases and 
rapid drops. Then a kind of chaotic region is followed. Above certain voltage, 
the total current behaves with a simple harmonic oscillation and the I-V 
characteristics form a multi-branch structure as in the current-biased case.  
The above behavior is the result of the inter-layer coupling, and may be used 
to confirm the inter-layer coupling mechanism of the formation of hysteresis 
branches.
\end{abstract}


\section{Introduction}
	I-V characteristics of the high T$_c$ superconductor 
Bi$_2$Sr$_2$Ca$_1$C$_2$O$_8$ (BSCCO) shows a strong hysteresis, producing 
multi-branches \cite{Muller1,Muller2,Muller3}. Many experiments have been 
reported to investigate properties of hysteresis branches 
\cite{Sakai1,Yurgens,Tanabe1,Itoh,Helm,Sakai2,Schlenga}, and it has turned 
out that there is rich physics in properties of I-V characteristics. 
Substructures are found in hysteresis branches, and are discussed in relation 
with phonon effects \cite{Helm}. The d-wave effect is also discussed in the 
higher voltage region, where the tunneling current becomes important 
\cite{Tanabe1,Schlenga}. 

It is well known in a single Josephson junction that the application of a 
constant voltage-bias leads a rather simple phenomenon of the harmonic 
current-oscillation with the frequency proportional to the applied voltage 
\cite{Tinkam}. This is due to the fact that the time-derivative of the phase 
difference is directly connected with the voltage, and that there is no room 
for the phase to move freely. Therefore, the most of the above mentioned 
experiments were performed with the current biased arrangement, aiming to 
investigate properties of I-V hysteresis branches, and, as long as we know, 
no systematic experiment with the voltage bias has been reported. 
We will point out in this paper that anomalous behaviors are 
expected, theoretically, in the low voltage-biased region in the case of 
the multi-layered intrinsic Josephson junctions of high T$_c$ superconductors, 
and that the voltage-biased experiment gives additional information 
on behaviors of I-V characteristics. 

In the multi-layered intrinsic Josephson junctions of high T$_c$ 
superconductors, inter-layer couplings among superconducting 
phase differences  become important due to the atomic scale of the thickness 
of the superconducting layer. In the ref. \cite{Koyama1} the mechanism of the 
inter-layer coupling is proposed as the charging effect of superconducting 
layers; the charging effect induces the variation of electric field and 
mediates the inter-layer coupling among superconducting phase differences. 
It has been shown that, by this mechanism, a branch structure in the I-V 
characteristics is obtained as the intrinsic nature of the strong anisotropy 
of high T$_c$ superconductors \cite{Machida}. 
In the previous paper \cite{Matsumoto1}, we have investigated the origin of 
hysteresis branches, and have shown that hysteresis jumps are induced when 
solutions change in nonlinear-coupled equations of phase-differences. 
Namely, the number of rotating phases and its patterns classifies solutions. 
At a point of a hysteresis jump, its number or distribution pattern is changed.

	The main difference between the cases of the single junction and the 
multi-junction lies in the fact that the applied voltage gives restriction 
only on the total voltage. Therefore, each junction may behave in a 
self-adjusted way, if an interaction effect allows it. In a simple array of 
$N$ single junctions, each junction feels only the $N$-th of the applied 
voltage, and one cannot expect much difference from the case of the single 
junction. However, in the intrinsic Josephson array of the high T$_c$ 
superconductors, the existence of an inter-layer coupling affects the 
collective motion of phase differences. Furthermore , the voltage control 
and current control lead different situations. Specially, in the voltage 
control case, one can reaches transient unstable states of the current biased 
case, since the applied voltage still enforces the total behavior of phase 
differences. In this paper, we investigate I-V characteristics in 
voltage-biased cases of high T$_c$ superconductors.  Since there is no
experiment available to refer at present, several cases with possible 
surface conditions will be presented. 

	In the next section, we summarize the formula for the voltage-biased 
case in the multi-Josephson junction model. In Sect. 3, we present numerically 
simulated results, and present I-V characteristics of the voltage-biased case. 
It will be shown that certain anomalous behavior of I-V characteristics 
appears in the applied low voltage region. We show that the voltage region is 
divided roughly into three regions, an anomalous periodic region, a chaotic 
region and a region of hysteresis branches. In an anomalous periodic region, 
the current shows a trigonometric functional increase and rapid drop. In a 
chaotic region, the total current behaves chaotically. In the last region, the 
behavior of the current is quasi-harmonic. Physical origin of these regions 
will be discussed. Sect. 4 is devoted to concluding remarks. 
\bigskip 

\def\dt{\frac{\partial}{\partial t}}
\section{Formula for the voltage biased multi-Josephson junction}
	In this section, we rewrite the formula for the multi-Josephson 
junction of ref. \cite{Matsumoto1} suitable to the voltage-biased analysis. 
The notations used here are same as those in ref. \cite{Matsumoto1}. 
Namely, we consider the $N+1$ superconducting layers, numbered from 
$0$ to $N$. We denote the gauge invariant phase difference of the $(l-1)$-th 
and $l$-th superconducting layer by $\varphi(l)$, and voltage by $V(l)$. 
The widths of the insulating and superconducting layers are denoted by $D$ and 
$s$, respectively. At the edges, the effective width of the superconducting 
layer may be extended due to the proximity effect into attached lead metals. 
The widths of the $0$-th and $N$-th superconducting layers are denoted by 
$s_0$ and $s_N$, respectively. 

In the voltage-biased case, the total voltage is equal to the applied external 
voltage $V_{ext}$,
\begin{equation}
	V_{ext}=\sum_{l=1}^NV_l \ . \label{Vcond}
\end{equation} 
In order to perform a numerical simulation under the condition (\ref{Vcond}), 
we rewrite the equation for the total current $J$ (Eq. (1) in 
ref. \cite{Matsumoto1}) and the equation relating the time-derivative of 
phase-difference and voltage (Eq. (3) in ref. \cite{Matsumoto1}) 
in the following way. First, we note that the total current $J$ is expressed 
by the external voltage $V_{ext}$ as 
\begin{equation}
	\frac{J}{J_c}=\frac{1}{N}\left(\sum_{l=l}^Nj_c(l)\sin(\varphi(l))
	+\beta \frac{V_{ext}}{V_p}
	+\frac{1}{\omega_p}\dt \frac{V_{ext}}{V_p}\right)\ . \label{JVext}
\end{equation}
The current $J$ is normalized by the critical current $J_c$, 
$j_c(l)=J_c(l)/J_c$  with $J_c(l)$ being the critical current for the $l$-th 
junction. The time $t$ is normalized by the inverse of the plasma frequency, 
$\omega _p$ = $\sqrt{\frac{2e}{\hbar}\frac{4\pi DJ_c}{\epsilon}}$, with 
$\epsilon$ being dielectric constant of the insulating layer. The voltage 
$V(l)$ is normalized by $V_p=\frac{\hbar\omega_p}{2e}$. The parameter $\beta$ 
is given by  $\beta=\frac{\sigma V_p}{J_cD}$. 
We define the voltage difference $v(l)$ as
\begin{equation}
	V(l)=\frac{1}{N}V_{ext}+v(l)\ \label{Vv}.
\end{equation}
Then we have the following coupled differential equations,
\begin{eqnarray}
	\frac{1}{\omega_p}\dt \frac{v(l)}{V_p}
	&=&-\beta \frac{v(l)}{V_p}-j_c(l)\sin(\varphi(l))
		+\frac{1}{N}\sum_{l'=1}^Nj_c(l')\sin(\varphi(l'))\label{dtv}\\ 
\noalign{\rm\noindent and }\nonumber\\ 
	\frac{1}{\omega_p}\dt \varphi(l)&=&\sum_{l'=1}^NA_{ll'}
	\left(\frac{1}{N}\frac{V_{ext}}{V_p}+\frac{v(l')}{V_p}\right)\ .
							\label{dtp}
\end{eqnarray}
A dissipation effect\cite{Matsumoto1,Ryndyk1,Ryndyk2}in superconducting 
layers will be neglected in this paper, and the matrix $A$ is given in 
Eq. (4) in ref. \cite{Matsumoto1}. 
It should be noted that Eq. (\ref{dtv}) gives
\begin{equation}
	\frac{1}{\omega_p}\dt \sum_{l=1}^Nv(l)=-\beta\sum_{l=1}^Nv(l)\ .
\end{equation}
Therefore if 
\begin{equation}
	\sum_{l=1}^Nv(l)=0 \label{cond}
\end{equation}
is satisfied at the initial time, it is satisfied in all time. 
In the next section, we present results of solving Eqs. (\ref{dtv}) and 
(\ref{dtp}) by numerical simulation. The restriction from the total voltage is 
achieved by putting the initial condition Eq. (\ref{cond}). 
The equation for $\varphi(l)$, Eq. (\ref{dtp}), indicates that the phase 
differences move according to $\frac{1}{N}V_{ext}$ as average. 
However, due to the mutual interaction, each $\varphi(l)$ shows a complicated 
behavior which affects the I-V characteristics and the behavior of the total 
current. 
\bigskip 

\section{Results of numerical simulation}
We have solved the coupled differential equations (\ref{dtv}) and (\ref{dtp}) 
by use of the fourth order Runge-Kutta method. The average current $J$ is 
obtained by
\begin{equation}
	J=\frac{1}{T_{max}-T_{min}}\int_{T_{min}}^{T_{max}}dtJ(t)\ .
\end{equation}
We choose the parameters as $N=10$, $\alpha=1.0$ and $\beta=0.2$, which has 
been used in ref. \cite{Matsumoto1}. 
In the simulation we have chosen the time step $\omega_pdt=1.0\times 10^{-3}$, 
$\beta\omega_pT_{min}=10.0$, $\beta\omega_pT_{max}=310.0$ and the voltage 
step $dV_{ext}=0.01$. 

	As Eq. (\ref{dtp}) shows, the phase $\varphi(l)$ increases according 
to $\int dt\frac{1}{N}V_{ext}$ when $V_{ext}$ is small. When $\varphi(l)$ 
reaches the value $\frac{\pi}{2}$, the phase goes into a rotating mode as can 
be seen from the analogy of the motion of a pendulum with a 
$\sin(\varphi(l))$-nonlinear term. This indicates that the whole motions of 
phases $\varphi(l)$ are much affected by the phase with the lowest $j_c$. 

	It may be difficult to arrange two electrode at edges exactly in 
the same condition. However, in order to show how surface conditions affect 
the behavior of I-V characteristics, we present the following three cases; 
1)$J_c(l)$  are symmetric but widths of surface are asymmetric, 
2) both $J_c(l)$ and widths of surface symmetric, and 3)$J_c(l)$ asymmetric 
and widths of surface symmetric. The case with $J_c(l)$ asymmetric and 
widths of surface asymmetric shows similar behaviors as the case 3). 
It should be noted that, even if the asymmetry between $s_0$ and 
$s_N$ is introduced, the observed critical current is $J_c$ in the 
current-biased case. 

We first assume that all $j_c(l)$ are equal,
\begin{equation}
	j_c(l)=1.0\ .
\end{equation}
and choose, for example, an asymmetric boundary condition 
$s_0/s=1.0$ and $s_N/s=2.0$. In the present analysis, it is enough to 
have slight difference between $s_0$ and $s_N$, the difference of proximity 
effect. 

	In Fig. 1(a), we show the overall behavior of I-V characteristics 
obtained from the adiabatic increase and decrease of the applied voltage 
(circle points) and those obtained from abrupt application of the voltage 
($\times$ points). We can see that the I-V characteristics show many 
hysteresis jumps, forming a branch structure. In the abrupt application of 
voltage, the data points in the low voltage region are scattered, although 
they lie on some linear I-V branches. Each I-V branch is classified by a 
number and pattern of rotating phases. When the voltage is large enough, all 
junctions have the rotating phase and the single I-V characteristic is 
obtained. In Fig. 1(a), we notice that there exists the region where the 
adiabatic and abrupt applications of voltage give the same results, in the 
low voltage region before the branch structure starts. 
In Fig. 1(b), we expand the low voltage region. We can identify 
the obvious branches labeled as I$_1$, I$_2$, II, III$_1$ and III$_2$. 
The crosses ($\times$) are for the abrupt voltage application. 
The lines I and II are same for three cases of the adiabatic increase, 
decrease and abrupt application of the voltage. The branches III$_1$ and 
III$_2$ are linear. The calculation shows that the I-V relations for III$_1$ 
and III$_2$ are same as those obtained from the hysteresis branches in the I-V 
characteristics of the current-biased case. On the line III$_1$, the phase 
$\varphi(1)$ is rotating and on the line III$_2$ the phases $\varphi(1)$ and 
$\varphi(N)$ are rotating. Other phases $\varphi(l)$ are oscillating. 
The cross points are grouped as one rotating-phase and two-rotating phase. 
The scattered points are identified by the distribution of the rotating-phases 
among $N$ junctions. 

	In Fig. 2, we show the I-V characteristics, when we choose a symmetric 
boundary condition $s_0/s=1.0$ and $s_N/s=1.0$. The branch II in Fig.1 is 
separated into II$_1$ and II$_2$. The branch III$_2$ is linear and is exactly 
same line as III$_2$ in Fig. 1. Although the branch II$_2$ is linear as 
average and lies on the same line as III$_1$, the calculated points are 
largely scattered, indicating a certain kind of instability. 
In order to obtain Fig. 2, we have to ensure the symmetry,
\begin{equation}
	\varphi(l)=\varphi(N+1-l)\ ,\ V(l)=V(N+1-l)\ ,
\end{equation}
in each time step by taking the average of $\varphi(l)$ and $\varphi(N+1-l)$, 
for example. Otherwise, a very small round error rapidly develops in the 
region II, and we obtained the similar result of Fig. 1(b), although we start 
with the symmetric boundary condition. 

	In order to see what kinds of solutions are realized in the regions I 
and II, we plot, in Fig. 3, the time-dependence of the phase $\varphi(l)$ and 
the current $J$ for the case of Fig. 2. Fig. 3(a),(b) are for the branches 
I$_1$ , and Fig. 3(c),(d) are for II$_1$. Fig. 3(b) shows periodic 
current behaviors, while Fig. 3(d) shows chaotic current behaviors. 
The results show the following. All $\varphi(l)$ increase slowly in the same 
way according to the applied voltage. When $\varphi(1)$ and $\varphi(N)$ at 
the edges reach the value $\frac{\pi}{2}$, they increase rapidly in the 
phase-rotating mode. Other $\varphi(l)$'s swing back. The phases $\varphi(1)$ 
and $\varphi(N)$ damp their motion by the effect of resistance $\beta$, and 
move slowly again in the same way as other phases. As the result, the current 
behaves periodically with trigonometric functional increase and rapid drop. 
The regions I$_1$ and I$_2$ are different in the values of phase rotation, 
$2\pi$ and $4\pi$, before damping. 
In Fig. 3(c), not only the phases at the edges but also other phases 
$\varphi(l)$ go into the rotating mode. Since oscillation and rotation 
of phases are interface in a complicated way, the chaotic current behavior 
is produced. The regions II$_1$ and II$_2$ are different in the values of 
phase rotation, $2\pi$ and $4\pi$, at the edges. 

	The above results lead us the following understanding of solutions of 
the multi-Josephson junction in the voltage-biased case. The applied voltage 
$V_{ext}$ forces the phase-differences $\varphi(l)$ of junctions to rotate. 
When some $\varphi(l)$ reaches $\frac{\pi}{2}$, it goes into the natural 
rotating mode with other phases swinging back. How much it rotates depends on 
its speed and resistance, forming new non-ohmic branches. When the rotation 
and oscillation of phase differences start to interact, solutions go into a 
chaotic region. When the voltage becomes large enough for certain phase 
difference to keep rotating, the I-V characteristics become ohmic and form 
multi-branches. 

	From the above understanding, we can immediately see that the I-V 
characteristics may change when there exists a certain junction with weaker 
$j_c$. Most likely, the surface layers may be damaged in the preparation 
process of electrodes. In fact in the previous analysis \cite{Matsumoto1}, 
we have to choose reduced $J_c$ values at the edges in order to fit the data. 
It was also reported that for thin film mesa structures the preparation 
process for the upper electrode can produce a cut of the first junctions, 
leading to an area about half size and thus half critical current for 
those junctions \cite{Seidel}. 
As an example, we choose 
\begin{equation}
	j_c(1)=0.5\ ,\ j_c(l)=1.0\ (l\neq 1)\ .
\end{equation}
The result is presented in Fig. 4. The chaotic region II is missing. 
In the inset Fig. 4(b), we show the enlarged figure of the region I. 
There are some drops and quasi-periodic structure in the I-V characteristics. 
Each part has a different value of $\varphi(l)$-phase rotation, $2\pi n$. 
We note that, in ref. \cite{Latyshev}, the appearance of certain periodicity 
in I-V characteristics has been reported, though it is for the sample 
of submicron junctions with no branch structures. It is a future problem 
to see if there is some relation with the present analysis. 

	We also point out that, in the analysis of Fig. 1, the regions I-II 
are obtained also when the external voltage is applied abruptly. 
Adiabatic and abrupt voltage application differ in the initial total energy. 
Therefore, the present result shows that the regions I-II have the single 
stationary state.
\bigskip

\section{Conclusion}
	In this paper, we have investigated the I-V characteristics of the 
voltage-biased case. Differently from the case of the single junction, 
I-V characteristics shows the characteristic behavior originated from the 
mutual inter-layer coupling mediated by the charging effect. Although each 
junction feels as average the external voltage as $\frac{1}{N}V_{ext}$, there 
is a still room for each phase to move self-consistently according to the 
interaction. Therefore, when the voltage is low, there is a competition of 
the voltage forced motion and the motion induced by the interactions. 
Adiabatic increase of the external voltage induces the following successive 
transition. In the low voltage, one has the region with a periodic oscillatory 
current of an asymmetric waveform, and then a possible region with a chaotic 
current oscillation follows. In these regions, non-ohmic branches are formed. 
In higher voltage, the region with a stable harmonic oscillatory current 
appears, having ohmic I-V branches. When there is a junction with a weaker 
$j_c$, the associated phase difference is easier to go into the rotating mode. 
Then the I-V characteristics have a quasi-periodic behavior in its non-ohmic 
region. The present analysis shows that voltage biased I-V characteristics 
reveals the importance of inter-layer coupling more strongly, and gives 
additional information to the current biased case. 
\bigskip 

\begin{center}
	{\Large\bf Acknowledgements}
\end{center}
	This work was supported by the Special Research Grant in the Faculty 
of Engineering, Seikei University (H. M. and S. S.), and Grant-in-Aid for 
Scientific Research (C) from Japan Society for the Promotion of Science 
(T. K.). 
\bigskip


%
%

\begin{figure}
	\includegraphics[15truecm, 18truecm]{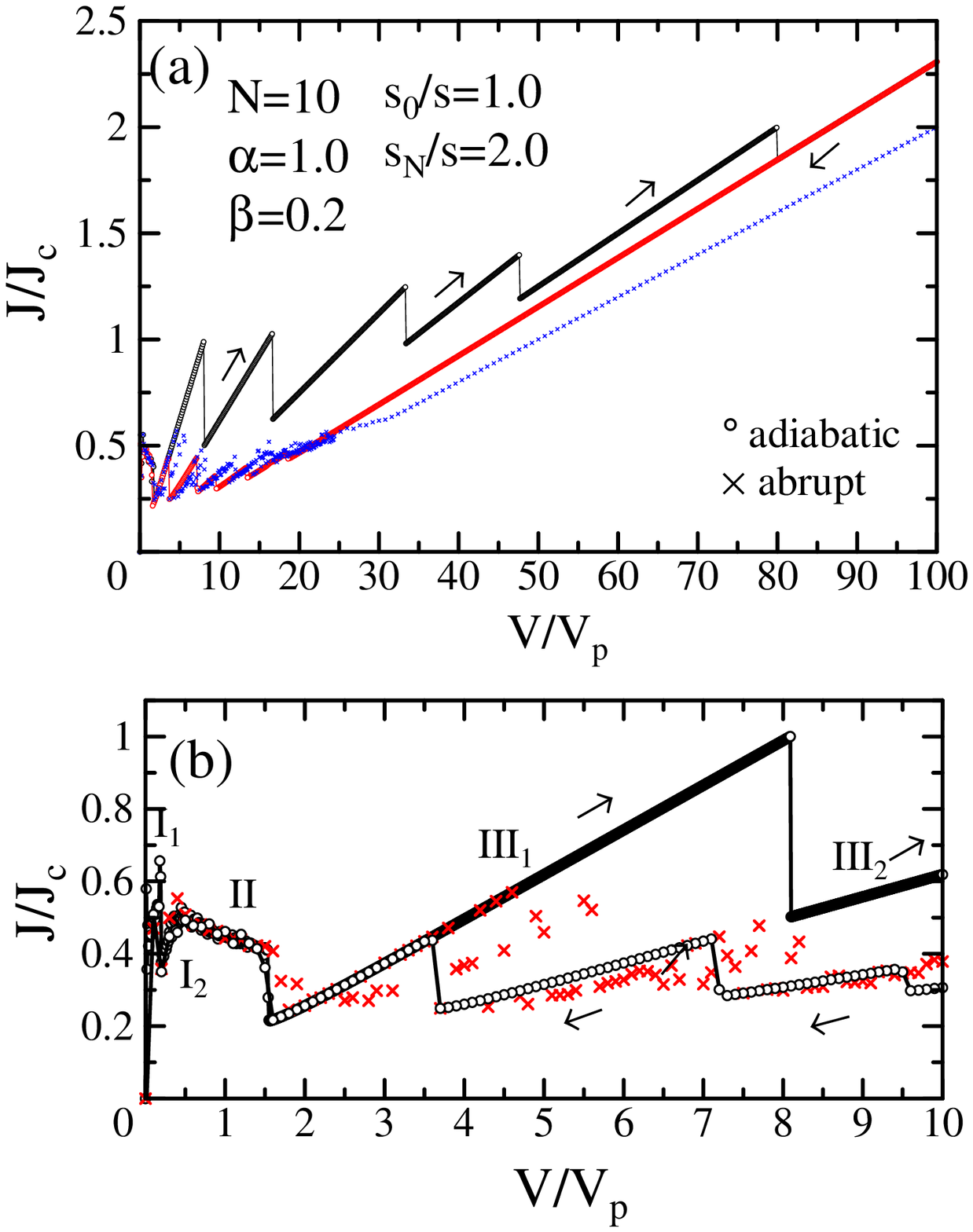} \bigskip 

\hspace{1.5truecm} \begin{minipage}{14truecm}
	Fig. 1. I-V characteristics with an asymmetric boundary condition\\ 
	 (a) for the wide range of $V$ and (b) for the expanded low voltage 
	region. The circles are for the adiabatic voltage change. 
	The crosses are for the abrupt voltage application. 
	Parameters are $\alpha=1.0$, $\beta=0.2$, $N=10$, $\frac{s_0}{s}=1.0$ 
	and $\frac{s_N}{s}=2.0$. 
        \end{minipage}
\end{figure} 
\newpage 
\begin{figure}
	\includegraphics[15truecm, 15truecm]{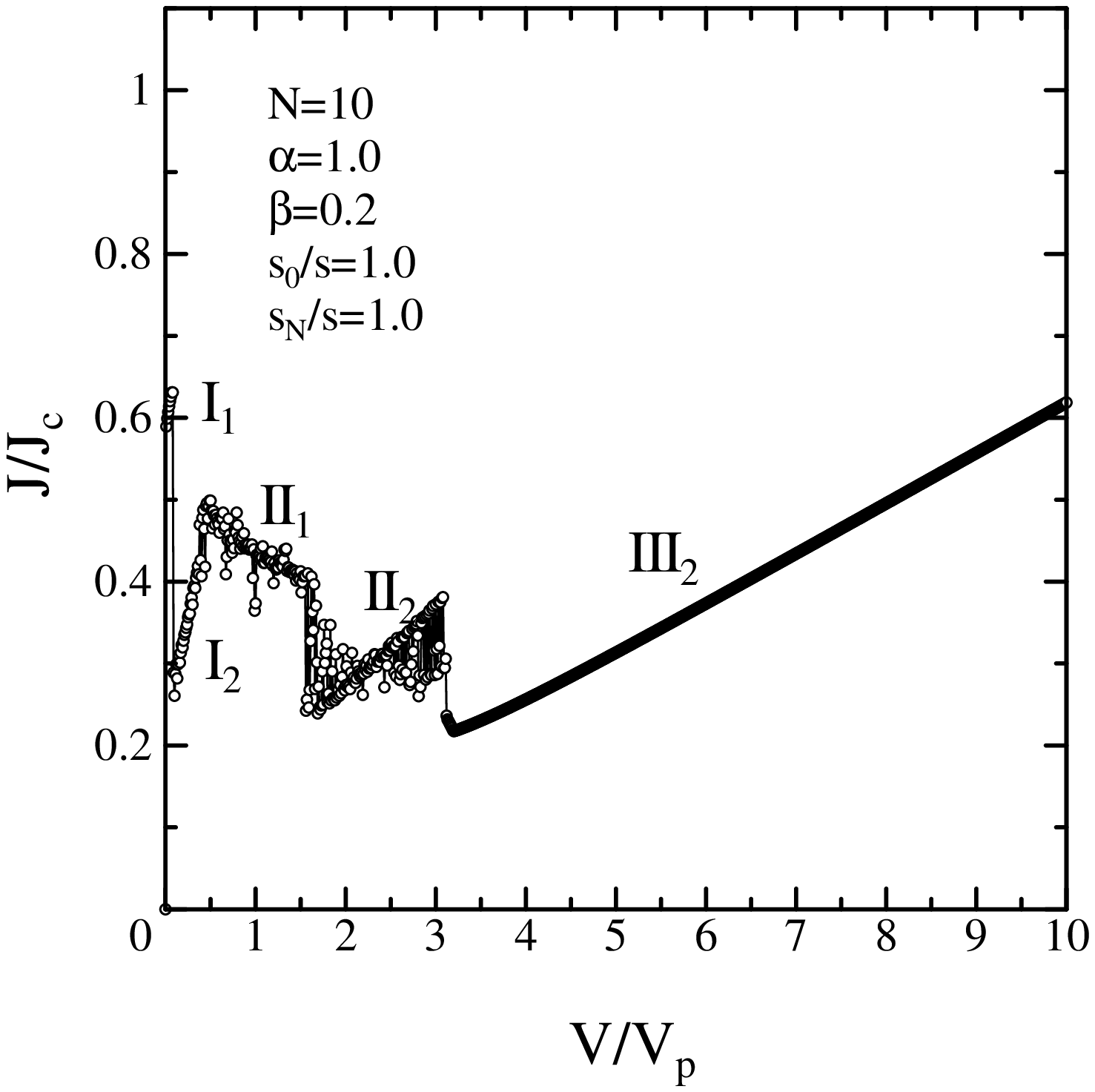} \bigskip 

\hspace{1.5truecm} \begin{minipage}{14truecm}
	Fig. 2. I-V characteristics with a symmetric boundary condition\\  
	Parameters are $\alpha=1.0$, $\beta=0.2$, $N=10$, $\frac{s_0}{s}=1.0$ 
	and  $\frac{s_N}{s}=1.0$. 
        \end{minipage}
\end{figure} 
\newpage 
\begin{figure}
	\includegraphics[15truecm, 18truecm]{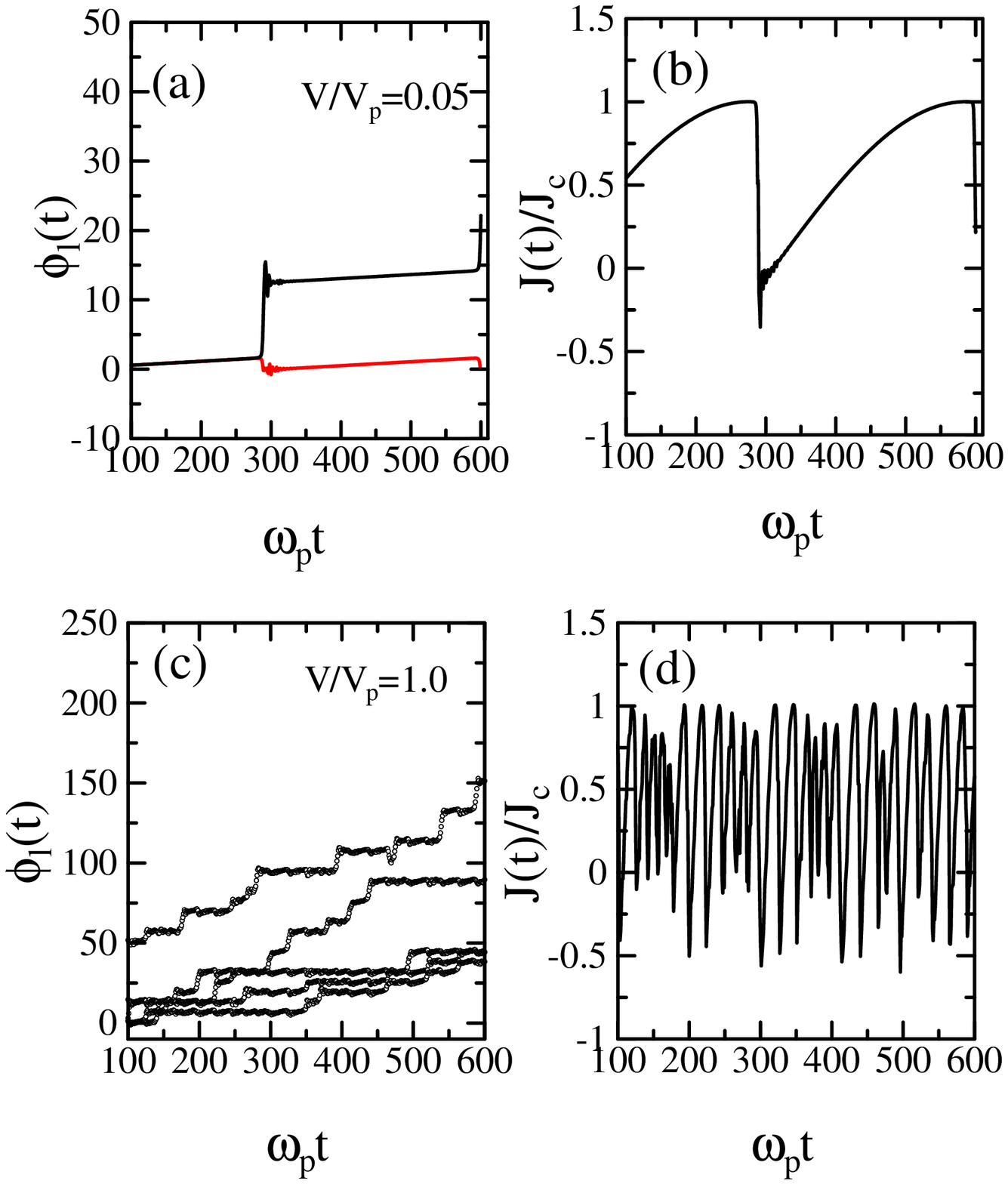} \bigskip 

\hspace{1.5truecm}\begin{minipage}{14truecm}
	Fig. 3. Time dependence of $\varphi(l)$ and $J/J_c$ for the branches 
	I$_1$ ( (a)-(b) ) and II$_1$ ( (c)-(d) )\\ 
	Parameters are $\alpha=1.0$, $\beta=0.2$, $N=10$, $\frac{s_0}{s}=1.0$ 
	and  $\frac{s_N}{s}=1.0$. 
        \end{minipage}
\end{figure} 
\newpage 
\begin{figure}
	\includegraphics[15truecm, 15truecm]{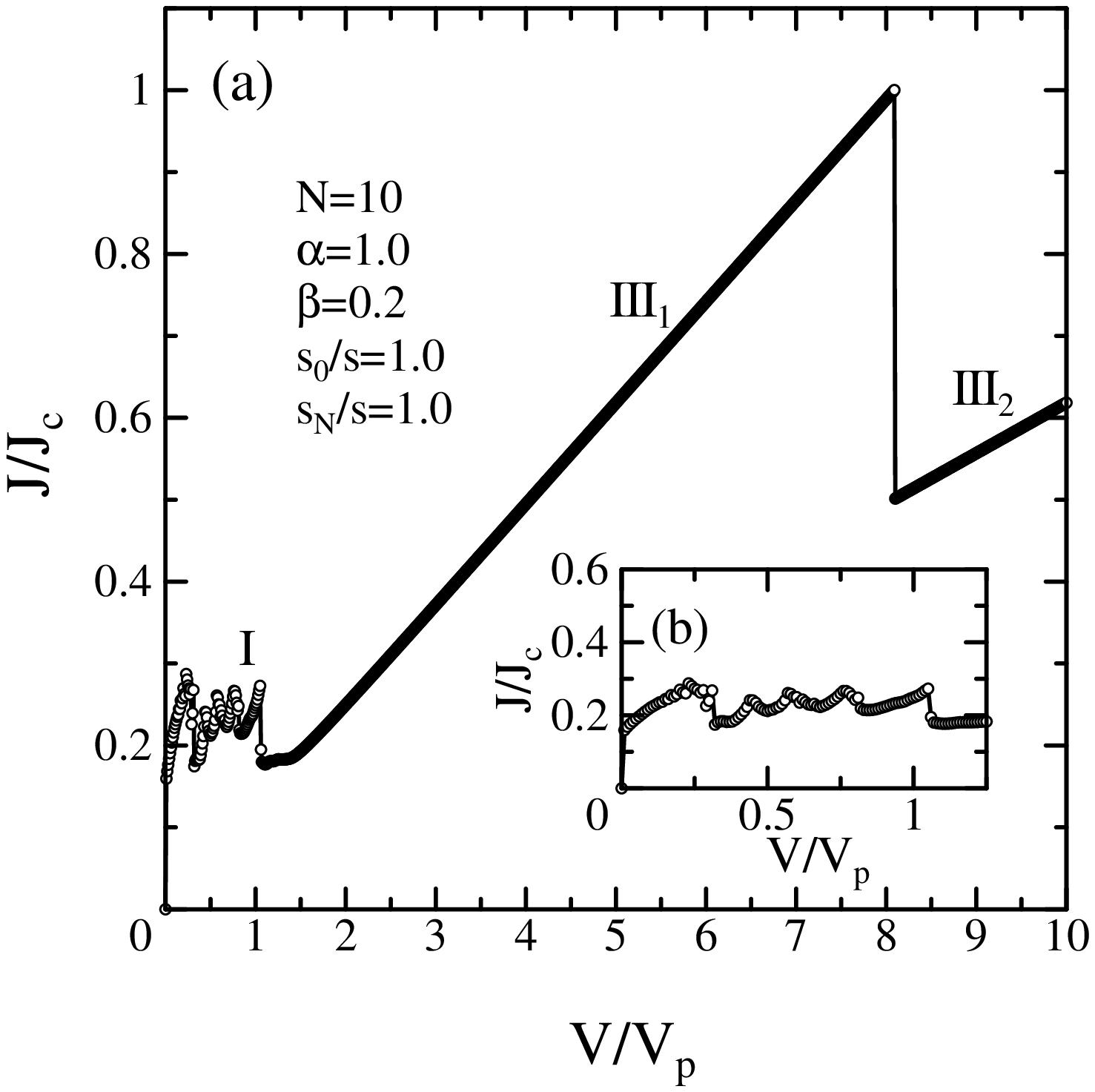}\bigskip
 
\hspace{1.5truecm}\begin{minipage}{14truecm}
	Fig. 4. I-V characteristics with a $j_c(l)$-asymmetric boundary 
	condition \\ 
	Parameters are $\alpha=1.0$, $\beta=0.2$, $N=10$, $\frac{s_0}{s}=1.0$, 
	$\frac{s_N}{s}=1.0$, $j_c(0)=0.5$ and $j_c(l)=1.0$ ($l\neq 0$). 
       \end{minipage}
\end{figure}

%
%

\end{document}